# Reorganization of the auditory-perceptual space across the human vocal range

Daniel Friedrichs, Volker Dellwo

Department of Computational Linguistics, University of Zurich
daniel.friedrichs@uzh.ch, volker.dellwo@uzh.ch

**ABSTRACT**

We analyzed the auditory-perceptual space across a substantial portion of the human vocal range (220–1046 Hz) using multidimensional scaling analysis of cochlea-scaled spectra from 250-ms vowel segments, initially studied in Friedrichs et al. (**2017**) J. Acoust. Soc. Am. **142** 1025–1033. The dataset comprised the vowels /i y e ø ɛ a o u/ (N=240) produced by three native German female speakers, encompassing a broad range of their respective voice frequency ranges. The initial study demonstrated that, during a closed-set identification task involving 21 listeners, the point vowels /i a u/ were significantly recognized at fundamental frequencies ($f_o$) nearing 1 kHz, whereas the recognition of other vowels decreased at higher pitches. Building on these findings, our study revealed systematic spectral shifts associated with vowel height and frontness as $f_o$ increased, with a notable clustering around /i a u/ above 523 Hz. These observations underscore the pivotal role of spectral shape in vowel perception, illustrating the reliance on acoustic anchors at higher pitches. Furthermore, this study sheds light on the quantal nature of these vowels and their potential impact on language evolution, offering a plausible explanation for their widespread presence in the world's languages.

## 1. INTRODUCTION

For much of the last century, it has been widely believed that vowels produced at high fundamental frequencies ($f_o$) are challenging to comprehend, supported by numerous studies in the field of singing research (e.g., [1,2]). In 2013, Sundberg [3] provided a comprehensive overview, summarizing the results from various studies in the field of Western classical singing, and concluded that vowel recognition declines at an $f_o$ of around 523 Hz (i.e., the musical note C5). Above this frequency, only the open vowels /a/ and /ɑ/ (i.e., the vowels with the highest $F_1$) remain identifiable, while identification rates for all other vowels drop towards chance level at higher $f_o$. This outcome is expected due to the sparse distribution of harmonics at high $f_o$, leading to an undersampling of the vocal tract transfer function and, consequently, a less precise specification of the formant frequency distribution, a crucial acoustic representation in the vowel identification process.

Contrary to these findings, a few early studies outside Western classical singing showed accurate vowel identification at high $f_o$ but did not consider secondary cues' impact [4,5]. Recent research has addressed these secondary cues at high $f_o$, such as vowel duration, formant transitions, and coarticulation (see [6,7,8] for more information on secondary cues). For example, Friedrichs et al. [9] found that the phonological function of the steady-state vowels /i y e ø ɛ a u o/ could be maintained at $f_o$ up to 880 Hz in a two-alternative forced-choice task. By increasing the number of response options and talker variability, another study [10] demonstrated that the point vowels /i a u/ could even remain identifiable up to 1046 Hz. These findings align with those reported by Zhang et al. [11], who discovered that isolated vowels /i a u/ produced by a female Chinese Yue Opera singer could be identified with high accuracy up to around 932 Hz when presented to phonetically trained listeners in a free-choice identification task.

Considering the conflicting observations in singing research, it is possible that vowel recognition at high $f_o$ is compromised by articulatory and acoustic adjustments made by classical Western Opera singers. Joliveau et al. [12] demonstrated, by measuring broadband acoustic excitation at the mouth of soprano singers, that they shift their first resonance frequency ($f_{R1}$) towards $f_o$ to increase vocal power in their performances. Such modifications were typically observed when $f_o$ was high and approached $f_{R1}$. Altering the vocal tract's first resonance frequency may, therefore, significantly impact the perceived vowel category.

Considering the diverse findings and the complex factors affecting vowel recognition at high $f_o$, this study aims to unravel the specific spectral characteristics of vowels ranging from 220 to 1046 Hz. Our goal is to discern the auditory-perceptual strategies [13] utilized across this broad vocal range, potentially illuminating the quantal nature of these vowels. This understanding could not only bridge the gap between the discrepancies observed in Western classical singing and other vocal techniques but also offer valuable insights for refining acoustic models for vowel recognition at high $f_o$.

## 2. METHODS

### 2.1. Participants

Twenty-one native German listeners participated, all students at the University of Zurich (10 females, 11 males; average age = 23.2 years, SD = 2.25). These participants were involved in a vowel identification task detailed in Friedrichs et al. [10]. All subjects had audiometric thresholds of less than 20 dB hearing level at octave frequencies between 125 and 8000 Hz. This task is distinct but complementary to an identification task undertaken during the corpus creation [14], where five phonetically trained listeners participated, whose performance forms a part of the discussion in this study.

### 2.2 Stimuli

Three female native German speakers (mean age = 32.2, standard deviation: 2.5) with professional vocal training in singing or acting were selected from a corpus of 70 speakers [14] based on their extended vocal range and similar $f_o$ in producing citation-form words and reading a story (mean 215.4 Hz, standard deviation 12.4 Hz). Recordings were made using a cardioid condenser microphone (Sennheiser MKH 40 P48) at a sampling frequency of 44.1 kHz and a constant distance of 30 cm from the speakers. The speakers produced the eight vowels /i y e ø ɛ a o u/ in isolation at ten target $f_o$ ranging from 220 to 1046 Hz (i.e., 220, 330, 440, 523, 587, 698, 784, 880, 988, and 1046 Hz). The target $f_o$ were played to the speakers as reference tones via studio monitor loudspeakers before each recording. $f_o$ was measured with an autocorrelation method [15,16] in Praat [17] and manually checked. For each vowel, multiple recordings were made, and those with the closest distance to the target $f_o$ (maximum deviation of 5%) were selected, resulting in a total of 240 vowels (10 $f_o$ x 8 vowels x 3 speakers). 250-ms segments were extracted from the vowel centers, normalized to an arbitrary intensity, and onset, and offsets were faded using raised cosines.

### 2.3. Procedure

In the listening experiment adopted from Friedrichs et al. [10], participants engaged in a mixed-talker listening task conducted in a noise-controlled environment at the University of Zurich. They were presented with vowel nuclei via Beyerdynamic DT 770 Pro, 250 Ω closed dynamic headphones. The participants viewed a screen displaying eight circular buttons, each denoting a vowel category arranged randomly, with a prompt "Welchen Vokal hörst Du?" (Which vowel do you hear?). Participants identified the presented vowel from these options, with the next stimulus auto-playing after a 1-second interval post-selection. Replay or repetition of a stimulus was not permitted, and each individual only encountered a particular vowel at each frequency once, amounting to 63 responses per vowel frequency option.

### 2.4. Cochlea-scaled spectra

Cochlea excitation patterns were simulated using a 200-channel linear gammatone filter bank, with bandwidths and center frequencies calculated using the ERB formulae by Glasberg and Moore [18]. The RMS level of the output wave was calculated and converted to dB for each filter channel. Frequency weighting was applied based on measurements made by Puria et al. [19] to account for the transmission properties of the middle ear.

### 2.5. Data analysis

Classical multidimensional scaling (MDS) analysis [20, 21] was performed on the cochlea-scaled spectra to model changes in the estimated auditory-perceptual space throughout the $f_o$ range. MDS has been previously used to assess perceptual proximity or dissimilarity in vowels [22, 23]. At each target $f_o$, all eight vowels were assigned to specific coordinates in a two-dimensional space derived from the calculated distances of the cochlea-scaled spectra. The distances shown in the MDS space are linearly related to the spectral distance.

Considering a vowel clustering phenomenon at high $f_o$ reported in previous studies [10,11], we aimed to investigate potential clustering of vowels at different $f_o$ by calculating the Euclidean distances between vowel pairs that appeared to cluster together in the MDS space (cluster 1: /i e y/, cluster 2: /ø ɛ a/, cluster 3: /u o/). For each speaker and $f_o$, we calculated the Euclidean distance between normalized cochlea-scaled spectra of all vowel pairs within each cluster using the formula:

$$(1) \quad d(p,q) = \sqrt{\sum (q_i - p_i)^2}$$

where $d(p,q)$ is the Euclidean distance between two spectra, and $p_i$ and $q_i$ are the amplitude values at frequency bin $i$ in spectra p and q, respectively. The summation is taken over all frequency bins $i$ from the first to the last frequency bin in the spectra.

To assess the relationship between the $f_o$ and the sum of Euclidean distances, we performed a piecewise mixed-effects linear regression with the Euclidean distances as the dependent variable and $f_o$ as a fixed effect. Based on Sundberg (2013), who suggests that listeners' vowel identification performance decreases at around 523 Hz, the model allowed for different slopes in the relationship between $f_o$ and the Euclidean distances, starting at 523 Hz. The random effects in this model included the speaker and listener. This analysis helped us determine if the changes in the Euclidean distances were significantly different across the $f_o$ range, providing insights into the clustering phenomenon and the underlying auditory-perceptual mechanisms involved throughout the human vocal range.

## 3. RESULTS

The results presented by Friedrichs et al. [10] demonstrated that listeners can identify the point vowels /i a u/ up to $f_o$ of nearly 1 kHz, although recognition rates for the vowels /y e/ decreased significantly, reaching chance levels for /e ø o/ at higher $f_o$. Interestingly, during the data corpus creation [14], five phonetically trained listeners were involved in an identification test, exhibiting an impressive accuracy rate of over 80% in distinguishing the selected vowels. Fig. 1 displays the estimated auditory-perceptual space, averaged across the three speakers at all recorded frequencies between 220 and 1046 Hz. At the lowest $f_o$ (220 Hz), the MDS space derived from cochlea-scaled spectra resembles a typical $F_1$–$F_2$ space. As the $f_o$ increases, the vowel height dimension (vertical axis) partially collapses while the frontness dimension (horizontal axis) expands, altering the ratio of these dimensions from 1:1.44 (220 Hz) to 1:0.47 (1046 Hz). Despite this reorganization, the point vowels /i a u/ remain at the corners of the vowel space.

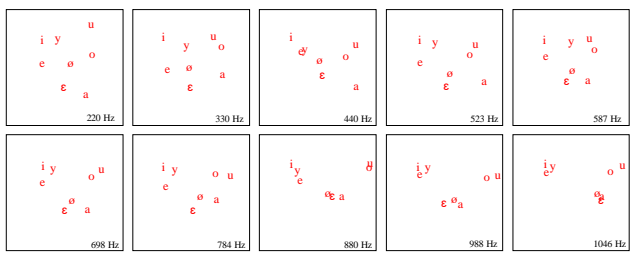

**Figure 1**: MDS plots illustrating the auditory-perceptual space derived from the cochlea-scaled spectra of vowels analyzed in this study, spanning the $f_o$ range of 220–1046 Hz (averaged and normalized across speakers). Beginning at approximately 523 Hz, the vowels commence clustering around the point vowels /i a u/.

Starting around 523 Hz, shifts towards the categories /i a u/ are observed: /y e/ move towards /i/, /o/ towards /u/, and /ø ɛ/ towards /a/. Fig. 2 illustrates the average decrease in Euclidean distance measurements for vowels in these three clusters as $f_o$ increases. Using the False Discovery Rate (FDR) controlling procedure [23], analyses of differences reveal no significant change in Euclidean distance measurements between the lowest $f_o$ (220 Hz) and 330 Hz (adjusted $p$ = 0.43) as well as 440 Hz (adjusted $p$ = 0.21), but significant differences occur between 220 Hz and 523 Hz (adjusted $p < .05$), as well as between 220 Hz and all $f_o$ up to 784 Hz (all adjusted $p < .05$) and up 1046 Hz (all adjusted $p < .001$).

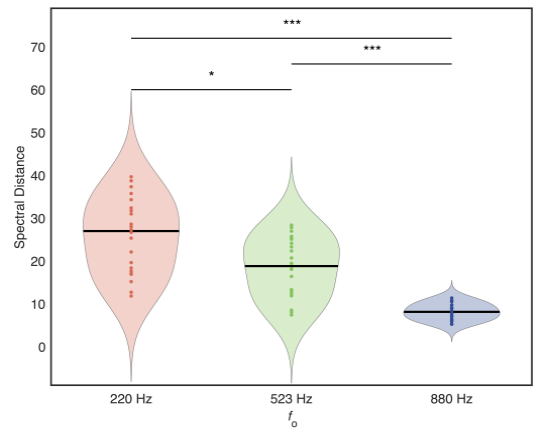

**Figure 2**: Violin plots showing the distribution of Euclidean distance measurements within the three clusters around the point vowels /i a u/ for a low, medium, and high $f_o$. The Euclidean distance between cochlea-scaled spectra decreases significantly towards the highest $f_o$.

Fig. 3 provides examples of cochlea-scaled spectra of /i y e/ forming a vowel cluster at high $f_o$. Dissimilarity is evident at the lowest $f_o$ (220 Hz), but spectral proximity becomes more apparent at higher $f_o$ (523 and 880 Hz). At 880 Hz, the Euclidean distance between the spectral vectors approaches the lower end of the scale and is only observable in high-frequency bands above 7 or 8 kHz.

Euclidean distances between the cochlea-scaled spectra of vowel pairs reveal a clear pattern across the investigated $f_o$. A piecewise linear mixed-effects regression with Euclidean distance as the dependent variable, $f_o$ as a fixed effect (split at 523 Hz), and speaker and listener as random effects showed no significant relationship between $f_o$ and Euclidean distance below 523 Hz (β1 = -0.005, $p$ = 0.65). At these lower $f_o$, the perceptual distance between vowels remains relatively constant. However, for $f_o$ above 523 Hz, a significant negative relationship emerges between $f_o$ and Euclidean distance (β2 = -0.045, $p < .001$), suggesting that vowel clustering becomes more pronounced at higher $f_o$, with vowels becoming increasingly perceptually similar.

At the lowest $f_o$ of 220 Hz, the median Euclidean distance was about 28 units, indicating a relatively large perceptual distance between vowels. As $f_o$ increased up to 523 Hz, the median Euclidean distance remained relatively stable. From 523 Hz onwards, the median Euclidean distance gradually decreased, reaching a minimum of about 8 units at 880 Hz.

The two higher $f_o$, 988 Hz and 1046 Hz exhibited nearly identical median Euclidean distances as observed at 880 Hz ($p > .05$), indicating that the vowel clustering phenomenon persists at these higher $f_o$. The piecewise linear mixed-effects regression also confirmed the lack of a significant difference in Euclidean distance between these three $f_o$ ($p > .05$). Overall, our results support the existence of a clustering phenomenon in vowel perception at higher $f_o$.

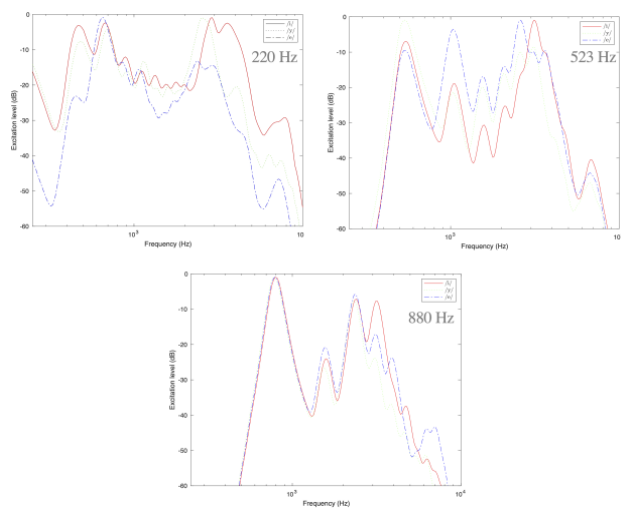

**Figure 3**: Cochlea-scaled spectra of the vowels /i y e/ produced by one speaker at fundamental frequencies of about 220, 523, and 880 Hz.

## 4. DISCUSSION

The findings presented in this study suggest that the auditory-perceptual space undergoes a systematic reorganization throughout the human vocal range and a clustering phenomenon around the vowels /i a u/ at very high $f_o$. The patterns found in the MDS analysis likely allow vowels to remain distinguishable even when the vocal tract transfer function is undersampled to a high degree. This might explain why studies on vowel perception outside the field of Western Opera singing reported vowel intelligibility at high $f_o$.

Our results indicate a tripartite organization of the auditory-perceptual space. At lower $f_o$ (220–440 Hz), the arrangement of the vowels resembles that of a vowel quadrilateral or a typical $F_1$–$F_2$ space. This suggests that well-established mechanisms, such as processing formant frequency distributions, remain unaffected in this range. An auditory-perceptual reorganization begins at frequencies around 523 Hz, previously identified as the absolute frequency at which vowel intelligibility drops for all except /a/-like vowels [3]. This reorganization process evolves until around 880 Hz. From this $f_o$ on (i.e., at 880, 988, and 1046 Hz), the auditory-perceptual clustering around the vowels /i a u/ is present, with three clusters formed by the vowels /i y e/, /ø ɛ a/, and /o u/.

While vowel category perception maintains a certain level of accuracy between 523–880 Hz, attributed to the distinctive spectral patterns visible in this frequency range, it seems to waver at the utmost $f_o$, given the spectral proximity within the clusters. Nonetheless, there seems to be a capacity for listeners to distinguish between vowels situated within the three clusters, even at the highest producible $f_o$. These observations not only resonate with the quantal theory of speech [25], which emphasizes the robustness in vowel production but also integrate well with the Natural Referent Vowel (NRV) framework [26]. The latter framework highlights the role of these vowels as referential anchors from the early stages of language acquisition, indicating their intrinsic robustness in perception from a young age. Consequently, our findings potentially offer insights into the pervasive presence of these vowels in languages worldwide, shedding light on the underlying reasons for their prevalence.

Several questions remain, however. The study garnered insights from an additional group of five phonetically trained listeners. They participated in a separate identification test with more than 80% correct responses for all selected vowels. Their expertise in phonetics and the opportunity to replay the stimulus possibly influenced the high-performance rate. In addition, it might also have aided their performance that they made judgments across over 30,000 vowels, which contributed to the creation of a larger corpus of vowel and voice studies [14,27,28]. Nevertheless, the contributing factors to their successful identification rates are unclear. This signals the necessity for further research to further investigate potential explanations.

Echoing the findings of Friedrichs et al. [10], and in line with results from previous studies utilizing vowels from the same corpus and other sources [11,29], our auditory-perceptual analysis, however, substantiates the important role of the point vowels at high $f_o$. The observed patterns highlight the critical role of whole spectrum features in vowel perception, suggesting that those may offer a fuller account than theories based solely on formant frequency patterns. The results also hint at the importance of acoustic anchors in the identification of other vowels, unveiling a potentially broader application and understanding of vowel perception at higher pitches.

## ACKNOWLEDGEMENTS

This study was supported by the NCCR Evolving Language, funded through the Swiss National Science Foundation Agreement #51NF40_180888. We are also grateful for the partial funding received from the SNSF Grants No. P2ZHP1_168375, P400PG_180693, and 100016_143943/1. We would like to express our gratitude to Dieter Maurer for his impactful work on vowel acoustics that inspired this research. Special thanks go to Stuart Rosen for his help with the auditory excitation patterns and to Paul Iverson for his contributions to the MDS analysis.

## 5. REFERENCES

[1] H. von Helmholtz, Die Lehre von den Tonempfindungen als Physiologische Grundlage für die Theorie der Musik. Vieweg+Teubner Verlag, 1913.

[2] H. Hollien, A. P. Mendes-Schwartz, and K. Nielsen, "Perceptual confusions of high-pitched sung vowels," J. Voice, vol. 14, pp. 287–298, 2000.

[3] J. Sundberg, "Perception of Singing," in The Psychology of Music, Elsevier Inc., 2013.

[4] D. Maurer, P. Mok, D. Friedrichs, and V. Dellwo, "Intelligibility of high-pitched vowel sounds in the singing and speaking of a female Cantonese Opera singer," in Proc. Annu. Conf. Int. Speech Commun. Assoc. INTERSPEECH, pp. 2132–2133, 2014.

[5] D. Maurer and T. Landis, "Intelligibility and spectral differences in high-pitched vowels," Folia Phoniatr. Logop., vol. 48, pp. 1–10, 1996.

[6] W. Strange, R. R. Verbrugge, D. P. Shankweiler, and T. R. Edman, "Consonant environment specifies vowel identity," J. Acoust. Soc. Am., vol. 60, pp. 213–224, 1976.

[7] D. Friedrichs, D. Maurer, H. Suter, and V. Dellwo, "Vowel identification at high fundamental frequencies in minimal pairs," in Proc. 18th Int. Congr. Phonetic Sci., paper no. 0438, pp. 1–5, 2015.

[8] D. Friedrichs, F. Nolan, and S. Rosen, "Do listeners rely on dynamic spectral properties in the recognition of high-pitched vowels?" in The Journal of the Acoustical Society of America, vol. 146, no. 4, p. 3051, 2019.

[9] D. Friedrichs, D. Maurer, and V. Dellwo, "The phonological function of vowels is maintained at fundamental frequencies up to 880 Hz," J. Acoust. Soc. Am., vol. 138, no. 1, pp. EL36–EL42, 2015.

[10] D. Friedrichs, D. Maurer, S. Rosen, and V. Dellwo, "Vowel recognition at fundamental frequencies up to 1 kHz reveals point vowels as acoustic landmarks," J. Acoust. Soc. Am., vol. 142, no. 2, pp. 1025-1033, 2017.

[11] Y. Zhang, F. Nolan, and D. Friedrichs, "Perceptual clustering of high-pitched vowels in Chinese Yue Opera," Speech Commun., vol. 137, pp. 60–69, 2022.

[12] E. Joliveau, J. Smith, and J. Wolfe, "Tuning of vocal tract resonance by sopranos," Nature, vol. 427, no. 6970, p. 116, 2004.

[13] D. Friedrichs, S. Rosen, P. Iverson, D. Maurer, and V. Dellwo, "Mapping vowel categories at high fundamental frequencies using multidimensional scaling of cochlea-scaled spectra," J. Acoust. Soc. Am., vol. 140, no. 4, p. 3219, 2016.

[14] D. Maurer, C. d'Heureuse, H., Suter, V., Dellwo, D., Friedrichs, T., Kathiresan, "The Zurich Corpus of Vowel and Voice Quality, Version 1.0," 2018.

[15] P. Boersma, "Accurate short-term analysis of the fundamental frequency and the harmonics-to-noise ratio of a sampled sound," Proc. Inst. Phonetic Sci., vol. 17, pp. 97–110, 1993.

[16] D. Friedrichs, D. Maurer, H. Suter, and V. Dellwo, "Methodological issues in the acoustic analysis of steady state vowels," in Trends in Phonetics and Phonology: Studies from German Speaking Europe, A. Leemann, M. J. Kolly, S. Schmid, and V. Dellwo, Eds., pp. 33–41, 2016.

[17] P. Boersma and D. Weenink, "Praat: doing phonetics by computer [computer program] (version 6.3.09)," 2023. [Online]. Available: http://www.praat.org/ (Last viewed March 22, 2023).

[18] B. R. Glasberg and B. C. J. Moore, "Derivation of auditory filter shapes from notched-noise data," Hear. Res., vol. 47, pp. 103–138, 1990.

[19] S. Puria, W. T. Peake, and J. J. Rosowski, "Sound-pressure measurements in the cochlear vestibule of human cadaver ears," J. Acoust. Soc. Am., vol. 101, pp. 2754–2770, 1997.

[20] R. N. Shepard, "The analysis of proximities: multidimensional scaling with an unknown distance function. I," Psychometrika, vol. 27, pp. 125-140, 1962.

[21] R. N. Shepard, "The analysis of proximities: multidimensional scaling with an unknown distance function. II," Psychometrika, vol. 27, pp. 219–246, 1962.

[22] P. Iverson and P. Kuhl, "Mapping the perceptual magnet effect for speech using signal detection theory and multidimensional scaling," J. Acoust. Soc. Am., vol. 97, no. 1, pp. 553–562, 1995.

[23] D. Kewley-Port and B. S. Atal, "Perceptual differences between vowels in a limited phonetic space," J. Acoust. Soc. Am., vol. 85, pp. 1726–1740, 1989.

[24] Y. Benjamini and Y. Hochberg, "Controlling the false discovery rate: A practical and powerful approach to multiple testing," J. Roy. Stat. Soc. Ser. B (Methodological), vol. 57, no. 1, pp. 289–300, 1995.

[25] K. N. Stevens, "On the quantal nature of speech," J. Phonetics, vol. 17, pp. 3–45, 1989.

[26] Polka, L., and Bohn, O.-S. (2003). Asymmetries in vowel perception. Speech Commun., 41(1), 221– 231.

[27] D. Maurer, H. Suter, D. Friedrichs, and V. Dellwo, "Acoustic characteristics of voice in music and straight theatre: Topics, conceptions, questions," in Trends in Phonetics and Phonology: Studies from German Speaking Europe, A. Leemann, M. J. Kolly, S. Schmid, and V. Dellwo, Eds., pp. 256–265, 2016.

[28] D. Maurer, Acoustics of the Vowel—Preliminaries. Bern, Switzerland: Peter Lang AG, International Academic Publishers, 2016.

[29] D. Friedrichs, "Beyond formants: vowel perception at high fundamental frequencies," Doctoral dissertation, University of Zurich, Zurich, Switzerland, 2017.